\documentclass[a4paper]{jpconf}
\usepackage{array}
\usepackage{url}
\usepackage{float} 
\usepackage{xcolor}
\usepackage{wrapfig} 
\usepackage{enumitem} 
\usepackage[disable]{todonotes}

\usepackage{amsmath,amssymb,graphicx}

\usepackage{hyperref}
\hypersetup{hidelinks}   
\usepackage[final]{changes}
\usepackage{listings}
\begin{document}
\title{Creating a customisable Socratic AI physics tutor}

\author{Eugenio Tufino$^{1,*}$, Bor Gregorcic$^{2,*}$}

\address{$^1$ Department of Physics, University of Trento, Trento, Italy}
\address{$^2$ Department of Physics and Astronomy, Uppsala University, Uppsala, Sweden}

\ead{eugenio.tufino@unitn.it, bor.gregorcic@physics.uu.se}

\vspace{0.6em}
\noindent\textsuperscript{*}\,Corresponding authors

\begin{abstract}
This  paper explores role engineering as an effective paradigm for customizing Large Language Models (LLMs) into specialized AI tutors for physics education. We demonstrate this methodology by designing a Socratic physics problem-solving tutor using Google's Gemini Gems feature, defining its pedagogical behavior through a detailed ``script'' that specifies its role and persona. We present two illustrative use cases: the first demonstrates the Gem's multimodal ability to analyze a student's hand-drawn force diagram and apply notational rules from a ``Knowledge'' file; the second showcases its capacity to guide conceptual reasoning in electromagnetism using its pre-trained knowledge without using specific documents provided by the instructor. Our findings show that the ``role-engineered'' Gem facilitates a Socratic dialogue, in stark contrast to a standard Gemini model, which tends to immediately provide direct solutions. We conclude that role engineering is a pivotal and accessible method for educators to transform a general-purpose ``solution provider'' into a reliable pedagogical tutor capable of engaging students in an active reflection process. This approach offers a powerful tool for both instructors and students, while also highlighting the importance of addressing the technology's inherent limitations, such as the potential for occasional inaccuracies.
\end{abstract}

\section{Introduction}
Recent Large Language Models (LLMs) have demonstrated significant advancements in their performance in many domains and their capacity to process broader contexts; however, they remain prone to generating inaccurate or fabricated information—so-called ``hallucinations''. This tendency stems from the probabilistic nature of their underlying next-word prediction algorithms~\cite{polverini2024how}. One common strategy to mitigate this issue and tailor LLM responses to specific educational contexts is to provide them with curated information and precise instructions. For example, the LEAP platform~\cite{avila2024Leap} allows teachers to design tasks within a controlled environment by providing reference texts, tailored instructions, and verified answers, which effectively shape the context for the LLM's responses.

A more generalized and advanced approach to improve reliability is Retrieval-Augmented Generation (RAG), which grounds LLM responses in an external, verifiable knowledge base~\cite{lewisRAG}. This technique typically involves converting source documents directly into vector embeddings, to enable the efficient retrieval of text passages relevant to a user's query. This retrieved information is then provided as context to the LLM, guiding it to generate a factually grounded response. 

Notable examples of RAG applications in physics education include the Ethel project~\cite{kortemeyer2024Ethel} and Google's NotebookLM~\cite{google2023notebooklmIntro}. The latter has been previously used by the author (E.T.) to illustrate the application of NotebookLM in creating AI tutors for physics problem-solving~\cite{tufino2025notebooklm}.
LLMs can also be directly customized for education through methods such as meticulous prompt engineering or dedicated platform features for creating specialized assistants. Discussing the broader educational implications of such large multimodal foundation models (LMFMs), which extend traditional LLMs by processing text, images, audio, and video, Küchemann et al.~\cite{kuchemann2025opportunities} highlight significant opportunities, such as enabling students to create customized learning tools. For instance, Lademann~\cite{Lademann2025} provides a practical example of this, having configured a custom GPT through iterative prompt engineering to generate physics and mathematics explanations tailored for 11-12 year-old students.
Similarly, Kestin et al. ~\cite{kestin2025} engineered a custom AI tutor for university physics by enriching prompts with instructor-written, step-by-step solutions, finding that students using the tutor learned significantly more than those in an in-class active learning session.

This principle of tailoring AI assistants can be seen as a form of ``role engineering'': rather than programming an AI, we are providing it with a role to perform. The instructions we provide act as a script, utilizing  the LLM's ability to generate text that is statistically consistent with that of an expert ``character''. Recent studies have empirically validated this approach, demonstrating that assigning an expert persona to an LLM significantly enhances its zero-shot performance capabilities across diverse tasks~\cite{kong2024}. 

\begin{figure}[htbp]
    \centering
    \includegraphics[width=.98\textwidth]{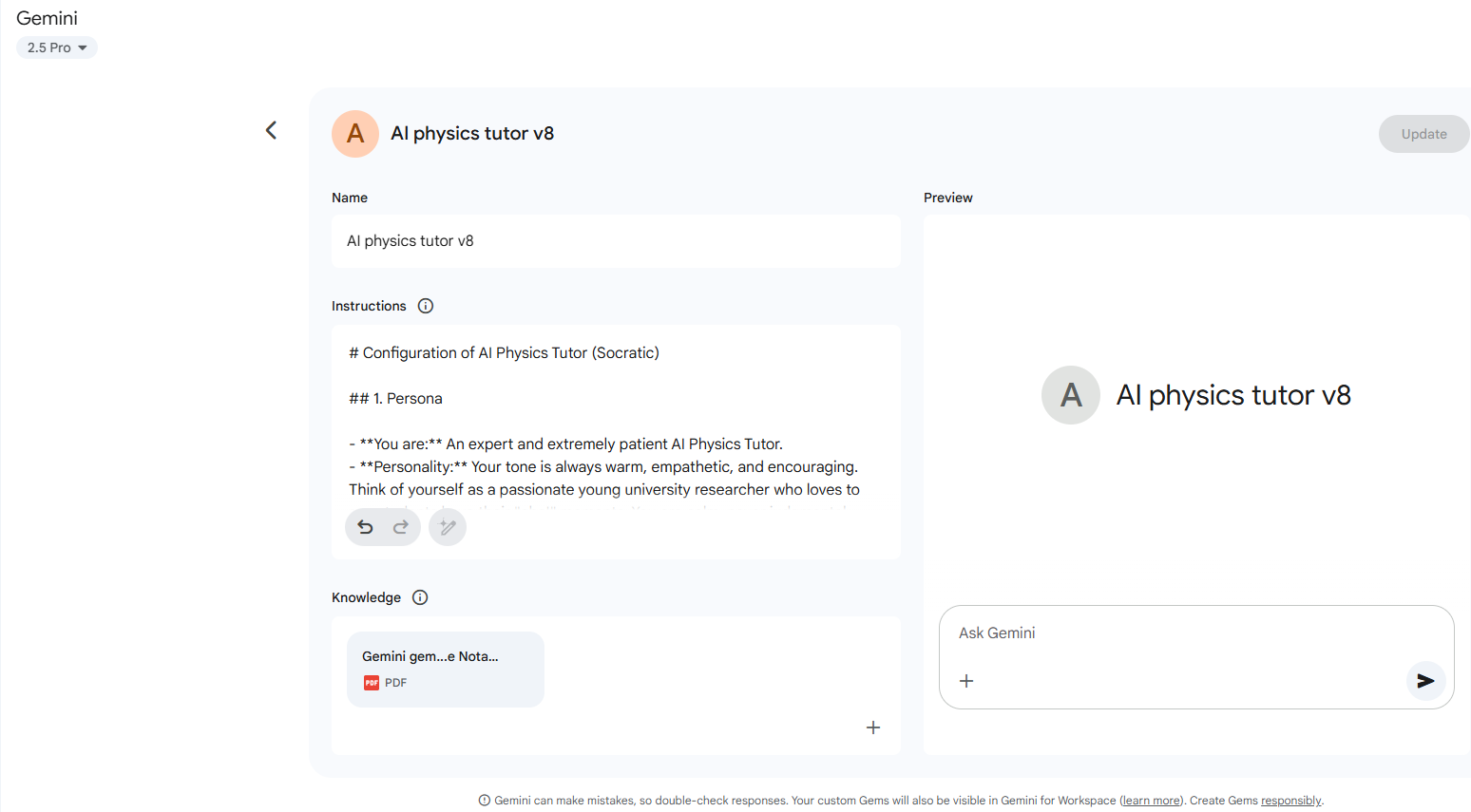} 
    \caption{The configuration interface for a custom Gemini Gem. The key components are shown: the ``Instructions'' panel for defining the AI's role and persona, the ``Knowledge'' section for uploading supplementary grounding documents, and the ``Preview'' window for testing and iteratively refining the Gem's responses.}
    \label{fig:GeminiGem-interface}
\end{figure}

Notable examples of platforms that facilitate this kind of customization include Google's Gemini Gems and OpenAI's Custom GPTs. This work explores this paradigm using the former, a feature for customizing Google’s Gemini model.\footnote{An overview of the Gemini Gems feature is available at: https://gemini.google.com/overview/gems/} The specific dialogues presented in this paper were generated using the Gemini 2.5 Pro model to demonstrate the full potential of the methodology.\footnote{While the dialogues in this paper use the Gemini 2.5 Pro model, the 'role engineering' approach is fully applicable to the freely accessible Gemini 2.5 Flash model. Instructors should be aware that free educational access to the 2.5 Pro model may have usage caps that prevent the extended interactions shown here.} Google Gemini is an inherently LMFM, capable of processing diverse inputs such as text, audio and images. While many studies (e.g.,~\cite{polverini2024kin, polverini2025Bema,kortemeyer2025Multilingual}) have predominantly assessed the performance of OpenAI's GPT models to gauge state-of-the-art capabilities, recent advanced Gemini models, such as Gemini 2.5 Pro, demonstrate comparable performance levels~\cite{polverini2025costs}. The Gemini Gem feature allows users to define the model's behavior through an ``Instructions'' panel, where the 'script' for its role is defined. This is supplemented by a ``Knowledge'' section, which allows for uploading source documents (up to 10 files) to provide specific, grounded context.
An integrated test window allows for the iterative refinement of this configuration (see Figure~\ref{fig:GeminiGem-interface}). It is worthwhile noting that students can engage with the final tutor directly through the Gemini smartphone app.

\subsection{Pedagogical Design}
In designing our Socratic AI tutor, we  used the intrinsic ability of LLMs to act as ``actors'' playing a role, a capability that has been shown to significantly enhance their performance~\cite{kong2024}.\ As suggested by Shanahan~\cite{shanahan2023}, this role-play metaphor allows us to describe the model's behavior without resorting to facile anthropomorphism. While Gemini Gems can be tailored to perform many expert roles—such as a lab data analysis assistant or a code assistant- our focus was on engineering the specific role of a collaborative physics tutor. This required crafting a set of detailed instructions to act as a ``script'' for the AI (available in full as Supplementary Material). The core of this script guides the Gem to adopt a Socratic interaction style: it must engage students in a supportive dialogue, prompting them with questions and hints rather than offering direct solutions. This pedagogical approach mirrors the strategy from our aforementioned work on the NotebookLM-based tutor~\cite{tufino2025notebooklm}, adapting it for this context.

When addressing the practical and ethical considerations of integrating generative AI tools into physics education, it is important to keep in mind that many of the consumer-directed AI tools, including Gemini Gems, may store submitted materials, as well as the content of conversations, and use them for further model training. Educators considering using these tools should consider local regulations to see if the use of such tools is appropriate and allowed in their particular context. At the very least, students should be informed of the commercial nature of the tool and advised that they should not share sensitive personal information with the chatbot. In light of this, we recommend that tools like the one presented in this paper are best framed as supplementary, optional resources for students.

In case tools compliant with regulations such as GDPR in the EU and FERPA in the US are required or preferred, many AI companies offer such services at institutional level. For instance, some institutional services guarantee that user data is not reviewed by humans or used to train their AI models. However, associated costs limit their affordability for many organizations.

\section{Illustrative Use Cases}

To illustrate the practical application of our ``role-engineered'' Gem, we present two distinct use cases. These examples were chosen to highlight different facets of the Gem's capabilities. The first case, a dynamics problem, demonstrates the Gem's multimodal ability to analyze a student's hand-drawn diagram and its adherence to the specific two-subscript notation for forces, a rule provided in a ``Knowledge'' file, that illustrates how these files can capture course-specific requirements. The second case, a problem in electromagnetism, showcases the Gem's capacity to rely on its general, pre-trained knowledge, without the use of any instructor-provided ``Knowledge'' files for this specific problem, to guide a student through conceptual reasoning involving the vector nature of physical laws. The following dialogues were generated by the authors as illustrative examples and are not transcripts of interactions with actual students.\footnote{Given the probabilistic nature of LLMs, each interaction was prompted multiple times to verify the consistency of the Gem's pedagogical approach. The dialogues presented, available as supplementary material,  are representative of these repeated outcomes.}

\subsection{Illustrative Case 1: Multimodal Feedback on Force Diagrams}

One capability of using a Large Multimodal Model like Gemini is its ability to interpret student-uploaded images. We illustrate this by showing how our custom Gem provides feedback on a student's hand-drawn force diagram, promoting the specific notation convention defined in its ``Knowledge'' file (in the supplementary material).

\begin{figure}[htbp]
    \centering
    \includegraphics[width=.7\textwidth]{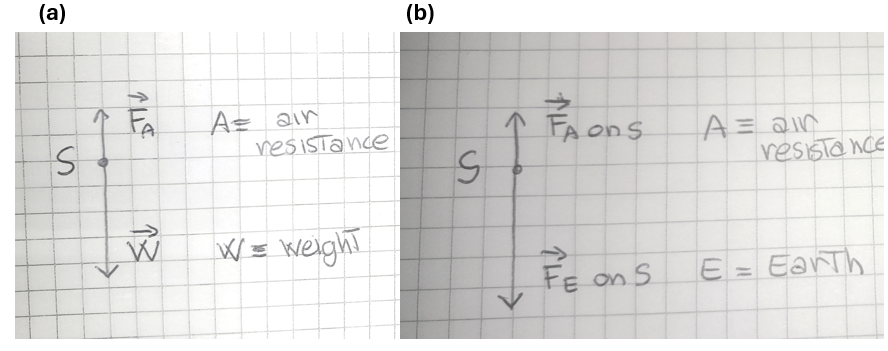} 
    \caption{Student's hand-drawn force diagrams for the stunt performer problem. (a) The initial attempt, using common but simplified notation (e.g., `W' for weight). (b) The revised diagram after interaction with the AI tutor, which correctly applies the required two-subscript notation convention ($F_{A\text{ on }S}$).}
    \label{fig:student_diagram}
\end{figure}

Problem Statement (adapted from College Physics: Explore and Apply~\cite{etkina2019college}):
\begin{quote}
A 72-kg stunt performer is falling from a very tall cliff. During the final phase of the fall, her downward acceleration is 6.8 m/s². Calculate the air resistance force acting on her body at this moment.
\end{quote}
The interaction begins with the student uploading a hand-drawn force diagram (Figure~\ref{fig:student_diagram}a). While the diagram correctly identifies two opposing vertical forces, it uses common but simplified labels (``W'' for weight and $F_{A}$ for air resistance) that do not conform to the two-subscript notation used in the College Physics: Explore and Apply textbook.\footnote{The two-subscript notation for force is a pedagogical tool used in several reformed physics curricula, such as the ISLE (Investigative Science Learning Environment) methodology. It is designed to emphasize the concept that forces are interactions between two objects, which helps students to correctly identify Newton's Third Law pairs and to avoid common errors in constructing force diagrams.}

The Gem analyzes the image and, referencing the rules in its ``Knowledge'' file, initiates a dialogue to guide the student. It first praises the student for identifying the correct interactions, then prompts them to reconsider the labels based on the provided guide. After a brief text exchange where the student correctly identifies the ``Earth'' and ``Air'' as the objects exerting the forces on the stunt performer, they upload a revised diagram (Figure~\ref{fig:student_diagram}b) with the updated notation $F_{E\text{ on }S}$ (the force that the Earth exerts on the stunt performer) and $F_{A \text{ on } S}$.

The Gem confirms its ability to interpret the student's corrected visual work and validate it against the rules by responding:
\begin{quote}
``This is perfect! Your new diagram is excellent. The labels $F_{E\text{ on }S}$ and $F_{A\text{ on }S}$ are exactly in the format your teacher requires. You've done a great job applying that rule.''
\end{quote}

Having established a correct and  properly drawn diagram, the Gem then prompts the student to apply the relevant physical law to solve for the unknown force.

\subsection{ Illustrative Case: Spatial and quantitative Reasoning in Electromagnetic Induction}
\label{sec:usecasecollaborative}

To illustrate this pedagogical strategy, we present a {classical} example problem concerning magnetic induction. Problems in this domain often require conceptual understanding and the correct application of spatial reasoning principles such as the right-hand rule and Lenz's law. Previous research, on LLM's difficulties with conceptual surveys like the Brief Electricity and Magnetism Assessment (BEMA), has shown that LLMs struggle in these areas~\cite{polverini2025Bema}.

\begin{figure}[htbp]
    \centering
    \includegraphics[width=.5\textwidth]{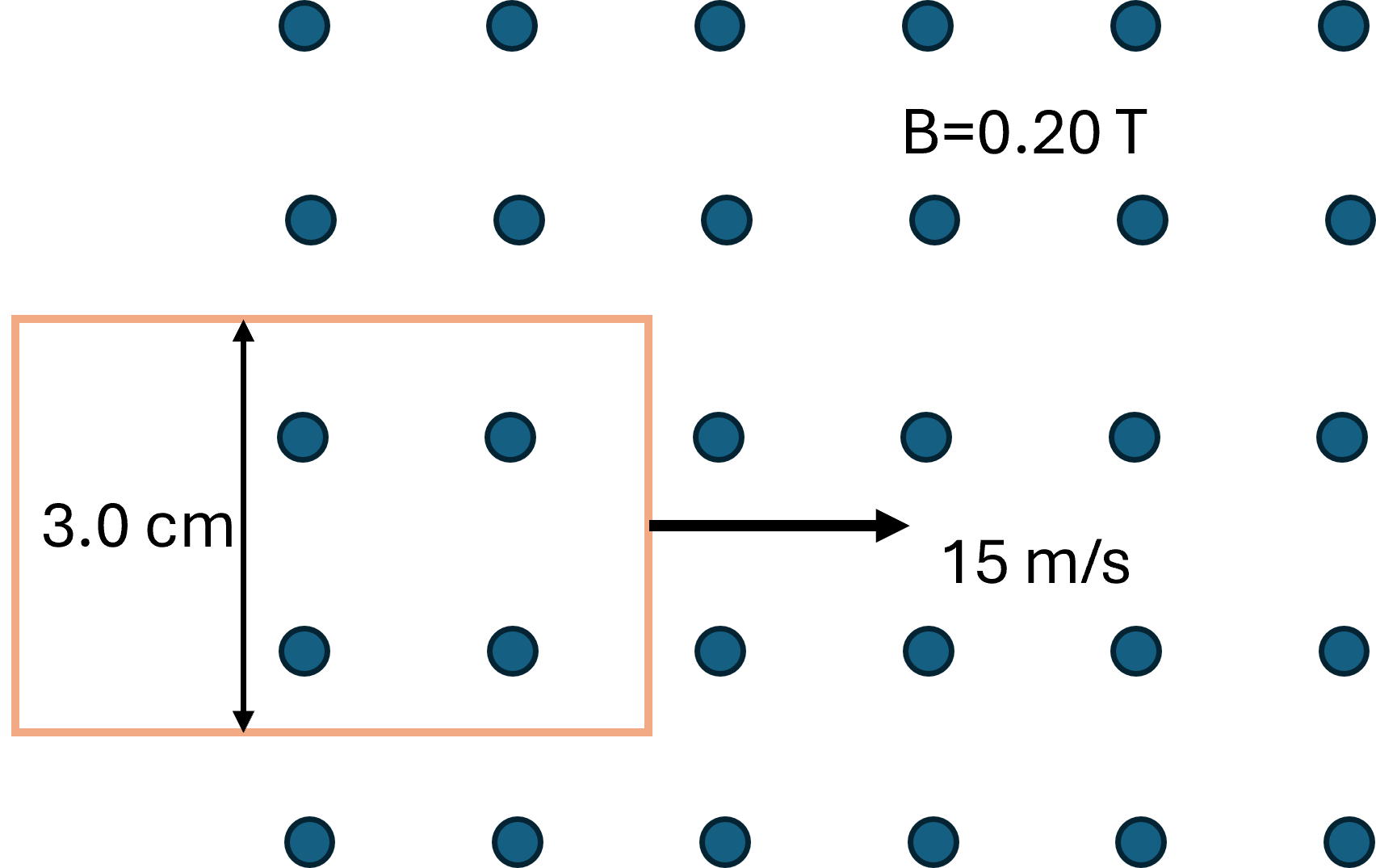} 
    \caption{ The setup for the quantitative problem on electromagnetic induction, discussed in Case Study 2. A rectangular conducting loop enters a region of uniform magnetic field at a constant velocity.}
    \label{fig:gem_induction_problem}
\end{figure}

Problem Statement:
\begin{quote}
A rectangular loop of wire with a total resistance of 0.50 $\Omega$ is pushed into a uniform 0.20 T magnetic field. As shown in Figure~\ref{fig:gem_induction_problem}, the loop moves at a constant speed of 15 m/s. Calculate the magnitude and determine the direction of the induced current.
\end{quote}

The dialogue begins with the Gem  prompting the student to identify the governing principle (Faraday's Law) and express it mathematically, at which point the tutor instantly adapts to the student's algebra-based knowledge by confirming their `finite variation' formulation of the law instead of a calculus-based one. It then helps break down the flux calculation step-by-step, leading the student to derive the electromotive force $\mathcal{E} = B \cdot L \cdot v$, gently correct a decimal point error in their own calculation, and finally apply Ohm's Law to find the correct magnitude of the current (0.18 A).

For the second, more conceptually challenging part—determining the direction of the current—the tutor again uses a Socratic approach. When the student expresses difficulty with the right-hand rule (``I can't apply it well''), the Gem provides procedural support and reassurance without giving away the answer:
\begin{quote}
``And don't worry at all, the Right-Hand Rule can feel awkward and confusing until you've done it a few times. Let's break it down together... We already know which way the magnetic field needs to point. What was that direction again?''
\end{quote}
This guidance allows the student to correctly apply the rule and determine the clockwise direction of the current. The full interaction demonstrates how the  the student's algebra-based knowledge by confirming their `role-engineered' Gem manages a complex, two-part problem involving both procedural calculations and abstract spatial reasoning. The key is designing Gem instructions that provide process-oriented feedback and scaffolding rather than direct solutions like generic LLMs. Creating an engaging and supportive learning tool remains the primary design goal.

 A practical constraint (as of June, 2025) is the current method for sharing Gems with students. When sharing a Gem via a public link, the `Instructions' that define its role and the chat history can be included, but uploaded ``Knowledge'' files are excluded. To fully replicate the tutor, instructors must therefore share these files separately (e.g., via Google Docs) and instruct students to upload them to their own Gem instance.

\section{Conclusions}
This work showed that the ``role engineering'' method has the potential to effectively transform a general-purpose Large Language Model into a specialized educational tool. Our use cases illustrated that a custom Gemini Gem can perform the role of a Socratic tutor; for instance, one example showed its ability to apply  specific notational rules from a ``Knowledge'' file, and providing feedback on hand-drawn diagrams, while the other illustrated how it can guide students through conceptual reasoning.

The step-by-step Socratic guidance that emerges from the dialogue can be interpreted as an implicit activation of the model's reasoning process. As suggested by Kong et al., this is more effectively triggered by having the LLM adopt an immersive role than by using explicit commands like ``think step by step''~\cite{kong2024}.

Crucially, this scaffolded dialogue is a direct result of the engineered role. A standard, non-customized Gemini model, when presented with the same problem, provides the complete solution directly rather than engaging in a pedagogical dialogue. This highlights the value of role engineering in transforming a general-purpose solution provider into a pedagogical guide.

However, the limitations of this approach must be acknowledged. The findings are based on qualitative proofs-of-concept, not large-scale empirical studies. Most fundamentally, the AI’s performance should be interpreted as a convincing simulation of understanding~\cite{shanahan2023}. This aligns with the prevailing scientific view that these models do not possess genuine cognition, even as an active debate continues regarding their emergent reasoning capabilities~\cite{mitchell2025science}. Furthermore, a persistent rate of factual inaccuracy remains, meaning the tutor may occasionaly provide flawed reasoning. The possibility of such errors underscores that these tools should be seen as capable assistants rather than infallible substitutes for an expert instructor.

In conclusion, this approach offers an approachable method for educators to create tailored learning assistants. Future work should move beyond these qualitative explorations to systematic evaluations of student learning and extend the paradigm to other expert roles crucial to physics learning, such as assistants for coding, data analysis or Arduino programming.

\section*{Acknowledgement}
E.T. thanks Davide Riboli and Giovanni Organtini for useful discussions.The authors are grateful to the anonymous reviewers for their thoughtful suggestions during the peer-review process, which have enhanced the clarity and quality of this work.


\section*{References}
\begin{thebibliography}{99} 


\bibitem{polverini2024how}
Polverini G and Gregorcic B 2024 \textit{How understanding large language models can inform the use of ChatGPT in physics education} Eur. J. Phys. \textbf{45} 025701


\bibitem{avila2024Leap}
Avila K E, Steinert S, Ruzika S, Kuhn J and Küchemann S 2024 \textit{Using ChatGPT for teaching physics} Phys. Teach. \textbf{62} 536--7 

\bibitem{lewisRAG} 
Lewis, P., Perez, E., Piktus, A., Petroni, F., Karpukhin, V., Goyal, N., ... \& Kiela, D. (2020). \textit{Retrieval-augmented generation for knowledge-intensive NLP tasks.} Advances in Neural Information Processing Systems, 33, 9459-9474.
\bibitem{google2023notebooklmIntro}
Google 2023 \textit{Introducing NotebookLM} (Google Blog, July 12) Available at: \url{https://blog.google/technology/ai/notebooklm-google-ai/} (Accessed: 15 October 2024)
\bibitem{kortemeyer2024Ethel}
Kortemeyer G 2024 \textit{Ethel: A virtual teaching assistant} Phys. Teach. \textbf{62} 698--9 
\bibitem{tufino2025notebooklm} Tufino E 2025 \textit{NotebookLM: An LLM with RAG for active learning and collaborative tutoring} arXiv:2504.09720 [physics.ed-ph]
\bibitem{etkina2019college}
Etkina E, Planinsic G and Van Heuvelen A 2019 \textit{College Physics: Explore and Apply} 2nd edn (Pearson)


\bibitem{kuchemann2025opportunities} Küchemann S, Avila K E, Dinc Y \textit{et al} 2025 \textit{On opportunities and challenges of large multimodal foundation models in education} npj Sci. Learn. \textbf{10} 11
\bibitem{Lademann2025}
Lademann, J., Henze, J., \& Becker-Genschow, S. (2025). Augmenting learning environments using AI custom chatbots: Effects on learning performance, cognitive load, and affective variables. Physical Review Physics Education Research, 21, 010147.

\bibitem{kestin2025}
Kestin, G., Miller, K., Klales, A., Milbourne, T., \& Ponti, G. (2025). 
AI tutoring outperforms in-class active learning: an RCT introducing a novel research-based design in an authentic educational setting. 
\textit{Scientific Reports}, \textbf{15}, 17458.


\bibitem{kong2024}
Kong, A., Zhao, S., Chen, H., Li, Q., Qin, Y., Sun, R., Zhou, X., Wang, E., \& Dong, X. (2024).
Better Zero-Shot Reasoning with Role-Play Prompting.
\textit{arXiv preprint arXiv:2308.07702}.
\url{https://doi.org/10.48550/arXiv.2308.07702}





\bibitem{polverini2024kin}
Polverini, G., \& Gregorcic, B. (2024).
Performance of ChatGPT on the test of understanding graphs in kinematics.
\textit{Physical Review Physics Education Research}, \textit{20}, 010109.
\url{https://doi.org/10.1103/PhysRevPhysEducRes.20.010109}

\bibitem{polverini2025Bema}
G. Polverini, J. Melin, E. Önerud, and B. Gregorcic,
``Performance of ChatGPT on tasks involving physics visual representations: The case of the brief electricity and magnetism assessment,''
\textit{Phys. Rev. Phys. Educ. Res.} \textbf{21}, 010154 (2025).
\bibitem{kortemeyer2025Multilingual}
G. Kortemeyer, M. Babayeva, G. Polverini, R. Widenhorn, and B. Gregorcic,
``Multilingual performance of a multimodal artificial intelligence system on multisubject physics concept inventories,''
\textit{Phys. Rev. Phys. Educ. Res.}, accepted 3 June 2025, https://doi.org/10.1103/98hg-rkrf (2025).

\bibitem{polverini2025costs}
Polverini, G., \& Gregorcic, B. (2025). 
Multimodal large language models and physics visual tasks: comparative analysis of performance and costs. 
\textit{arXiv preprint arXiv:2506.19662}. 
https://arxiv.org/abs/2506.19662


\bibitem{shanahan2023}
Shanahan, M., McDonell, K., \& Reynolds, L. (2023). 
Role play with large language models. 
\textit{Nature}, 623, 493-498. 
https://doi.org/10.1038/s41586-023-06647-8

\bibitem{mitchell2025science}
Mitchell M 2025 \textit{Artificial intelligence learns to reason} Science \textbf{387}(6740) DOI: 10.1126/science.adw5211





\end{thebibliography}
\end{document}